\documentclass[a4paper,fleqn,usenatbib]{mnras}
\usepackage{newtxtext,newtxmath}
\usepackage[T1]{fontenc}
\usepackage{ae,aecompl}
\pdfoutput=1
\usepackage{graphicx}	
\usepackage{amsmath}	
\usepackage{multicol}  
\usepackage{threeparttable}

\title[Molecular gas in the halo of TXS 0828+193]{ALMA detects molecular gas in the halo of the powerful radio galaxy TXS 0828+193}
\author[J. Fogasy et al.]{
Judit Fogasy,$^{1}$\thanks{E-mail: judit.fogasy@chalmers.se}
K. K. Knudsen,$^{1}$
G. Drouart,$^{2}$
and B. Gullberg,$^1$
\\
$^{1}$Department of Space, Earth and Environment, Chalmers University of Technology, Onsala Space Observatory, 439 92 Onsala, Sweden\\
$^{2}$Curtin University, Department of Physics and Astronomy, Kent Street, Bentley, 6102, Perth, WA, Australia\\
}

\date{Accepted XXX. Received YYY; in original form ZZZ}

\pubyear{2020}

\begin{document}
\label{firstpage}
\pagerange{\pageref{firstpage}--\pageref{lastpage}}
\maketitle

\begin{abstract}
Both theoretical and observational results suggest that high-redshift radio galaxies (HzRGs) inhabit overdense regions of the universe and might be the progenitors of local, massive galaxies residing in the centre of galaxy clusters.
In this paper we present CO(3--2) line observations of the HzRG TXS 0828+193 ($z=2.57$) and its environment using the Atacama Large Millimeter/submillimeter Array. In contrast to previous observations, we detect CO emission associated with the HzRG and derive a molecular gas mass of $(0.9\pm0.3)\times10^{10}\,\rm M_{\odot}$. Moreover, we confirm the presence of a previously detected off-source CO emitting region (companion \#1), and detect three new potential companions. The molecular gas mass of each companion is comparable to that of the HzRG. Companion \#1 is aligned with the axis of the radio jet and has stellar emission detected by \textit{Spitzer}. Thus this source might be a normal star-forming galaxy or alternatively a result of jet-induced star formation. The newly found CO sources do not have counterparts in any other observing band and could be high-density clouds in the halo of TXS 0828+193 and thus potentially linked to the large-scale filamentary structure of the cosmic web.
\end{abstract}




\section{Introduction}

Powerful high-redshift radio galaxies (HzRGs) are among the favoured targets to study the evolution of massive local galaxies and understand the correlations found between the supermassive black hole (SMBH) mass and the bulge mass or velocity dispersion of ellipticals \citep[e.g.][]{1998AJ....115.2285M, 2000ApJ...539L..13G, 2001ApJ...547..140M, 2002ApJ...574..740T, 2003ApJ...589L..21M, 2004ApJ...604L..89H, 2013ApJ...764..184M}. HzRGs harbour actively accreting SMBHs and also show signs of intense star formation via luminous sub-millimetre (submm) emission (see \citealt{2008A&ARv..15...67M} for a review). 
Observations of HzRGs found that their SMBHs can become overly massive compared to their host galaxies if the timescale of black hole accretion is similar to that of the star formation \citep{2019A&A...621A..27F}.
In order for these galaxies to be consistent with the observed local relations, apart from in situ star formation other processes, minor mergers in particular, might be the key to grow their stellar mass \citep{2014A&A...566A..53D, 2019A&A...621A..27F}. 
Indeed, HzRGs are often located in overdense regions, possibly protoclusters, surrounded by companion galaxies both at large and small scales \citep[e.g.][]{2007A&A...461..823V, 2010MNRAS.405..347F, 2011MNRAS.410.1537H, 2011MNRAS.417.1088K, 2012ApJ...749..169G, 2013A&A...559A...2G, 2013ApJ...769...79W, 2016ApJ...830...90N}.
Thus, the detailed study of HzRGs and their environment through submm continuum and CO emission line observations is essential to understand the ongoing processes shaping their evolution.\\
\indent
In this paper we focus on the HzRG TXS 0828+193 ($z=2.57$) and its large-scale environment.
TXS 0828+193 is a Fanaroff-Riley class II radio galaxy, with extended radio lobes ($\sim$ 100 kpc) and multiple hotspots \citep{1994A&AS..108...79R, 1997ApJS..109....1C}. Rest-frame UV observations show clumpy and asymmetric morphology \citep{1999A&A...341..329P}. The HzRG has an extended ($\sim$ 80 kpc), kinematically unperturbed gas reservoir traced by $\rm Ly\alpha$ emission \citep{2002MNRAS.336..436V}. Rest-frame optical spectroscopic observations revealed that the radio galaxy is a luminous line emitter,
 indicating the presence of a strong galactic outflow \citep{2008A&A...491..407N}. The infrared spectral energy distribution of the radio galaxy has been studied in detail, using \textit{Spitzer} and \textit{Herschel} observations \citep{2010ApJ...725...36D,2014A&A...566A..53D, 2016A&A...593A.109D}.
Based on \textit{Spitzer} IRAC (Infrared Array Camera) observations used in the Clusters Around Radio-Loud AGN (CARLA) program, the large-scale environment of TXS 0828+193 is overdense at $\geq3\sigma$ level, implying that it resides in a rich galaxy cluster \citep{2013ApJ...769...79W}.

\citet{2009MNRAS.395L..16N} studied the molecular gas of TXS 0828+193 through observation of the CO(3--2) line using the IRAM PdBI. These observations
did not detect CO emission associated with the HzRG, but revealed a luminous CO(3--2) line emitting region aligned with the axis of the radio jet, $\sim$10 arcsec ($\sim$80 kpc) south-west of the radio galaxy. The CO emission is not resolved within the 5 arcsec beam size of the observation, which suggests an upper limit on the size of $\sim40$ kpc. However, the integrated CO spectrum shows two peaks, denoted as SW1 and SW2, with a velocity offset of $-200\pm40\,\rm km\,s^{-1}$ ($z_{\rm sw1}=2.5761\pm0.0008$) and $-920\pm70\,\rm km\,s^{-1}$ ($z_{\rm SW2}=2.5678\pm0.0007$) relative to the HzRG. This suggests two components within the same spatially unresolved component. The total molecular gas mass of the two components is $\rm M_{\rm H_2} \sim 1.4\times10^{10}$\,M$_\odot$. The CO emission region does not have a counterpart within the 5 arcsec beam in the shallow \textit{Spitzer} IRAC observation (PI: Nesvadba, program ID 60147) and MIPS 24 $\mu$m bands, nor in the optical. 

\citet{2009MNRAS.395L..16N} proposed two possible scenarios for the origin of the CO emission. The CO emission might be associated with a very young, gas-rich companion galaxy with a stellar mass of $<3\times10^9$ M$_\odot$, hence the non-detection in the mid-IR bands. Alternatively, the source of the CO emission is a cold cloud or filament in the halo of the HzRG. 

With the aim of revealing the origin of the luminous CO emitting region, we observed the HzRG with the Atacama Large Millimeter/submillimeter Array (ALMA), obtaining CO(3--2) and continuum measurements. As HzRGs are known for residing in overdense environment, we also search for additional companion galaxies. In section \ref{sec:observations} we present the ALMA observations, in section \ref{sec:analysis} we describe the results of the observations and the data analysis, including the search for additional CO sources in the field. In section \ref{sec:discussion} we discuss our findings and compare them to other studies found in the literature. Finally, we summarise our results in section \ref{sec:conclusion}.
Throughout this paper we adopt WMAP7 cosmology with $H_{0}=70.4\ \rm{km \ s^{-1} Mpc^{-1}}$, $\rm \Omega_{m}=0.272$ and $\rm \Omega_{\Lambda}=0.728$ \citep{2011ApJS..192...18K}.

\section{Observations}
\label{sec:observations}
TXS 0828+193 was observed with ALMA during cycle 3 on 2016 June 3 and June 11  using 42 and 36 antennae (project code 2015.1.00661.S; PI: Fogasy). The on-source integration time was 148.2 min. One of the  1.875 GHz spectral windows was tuned to the central frequency of 96.818 GHz to cover the redshifted CO(3--2) line, while the other three spectral windows were used to observe the continuum. 
The data calibration was done applying the ALMA Science Pipeline, which includes standard calibration and reduction steps, such as flagging, bandpass calibration,  flux and gain calibration. The quasar J0854+2006 was used as a bandpass calibrator source.
We assume a conservative 10 percent absolute flux calibration error.

The imaging of data was carried out with the Common Astronomy Software Applications (\textsc{casa} v.5.4.0; \citealt{2007ASPC..376..127M}), using the \textsc{tclean} algorithm using natural weighting. The continuum was imaged excluding the spectral window containing line emission ($96.53-97.17$ GHz) and cleaned to the threshold of $0.01\,\rm mJy$. The resulting continuum image has an rms of $6.9\, \mu\rm Jy\,beam^{-1}$ at the centre of the primary beam and a beam size of $1.53\arcsec\times1.33\arcsec$ with a P.A. of $-48.02\degr$. 

The continuum was fitted using the line free channels of every spectral window and was subtracted from the spectral line data in the $uv$-plane using the \textsc{uvcontsub} task in \textsc{casa}.
The line data was imaged with  a width of 6 channels, resulting in a velocity binning of  $72.6\,\rm km\,s^{-1}$ and has an rms of $(0.11-0.12)\,\rm\,mJy\,beam^{-1}$ at the centre of the primary beam. The median area beam of the line data is $1.59\arcsec\times1.40\arcsec$ with a P.A. of $-44.37\degr$. For both the continuum image and line data cube we applied primary beam correction.

\section{Results and analysis}
\label{sec:analysis}

\subsection{ALMA continuum emission}

\begin{figure}
	\centering
	\includegraphics[width=\columnwidth]{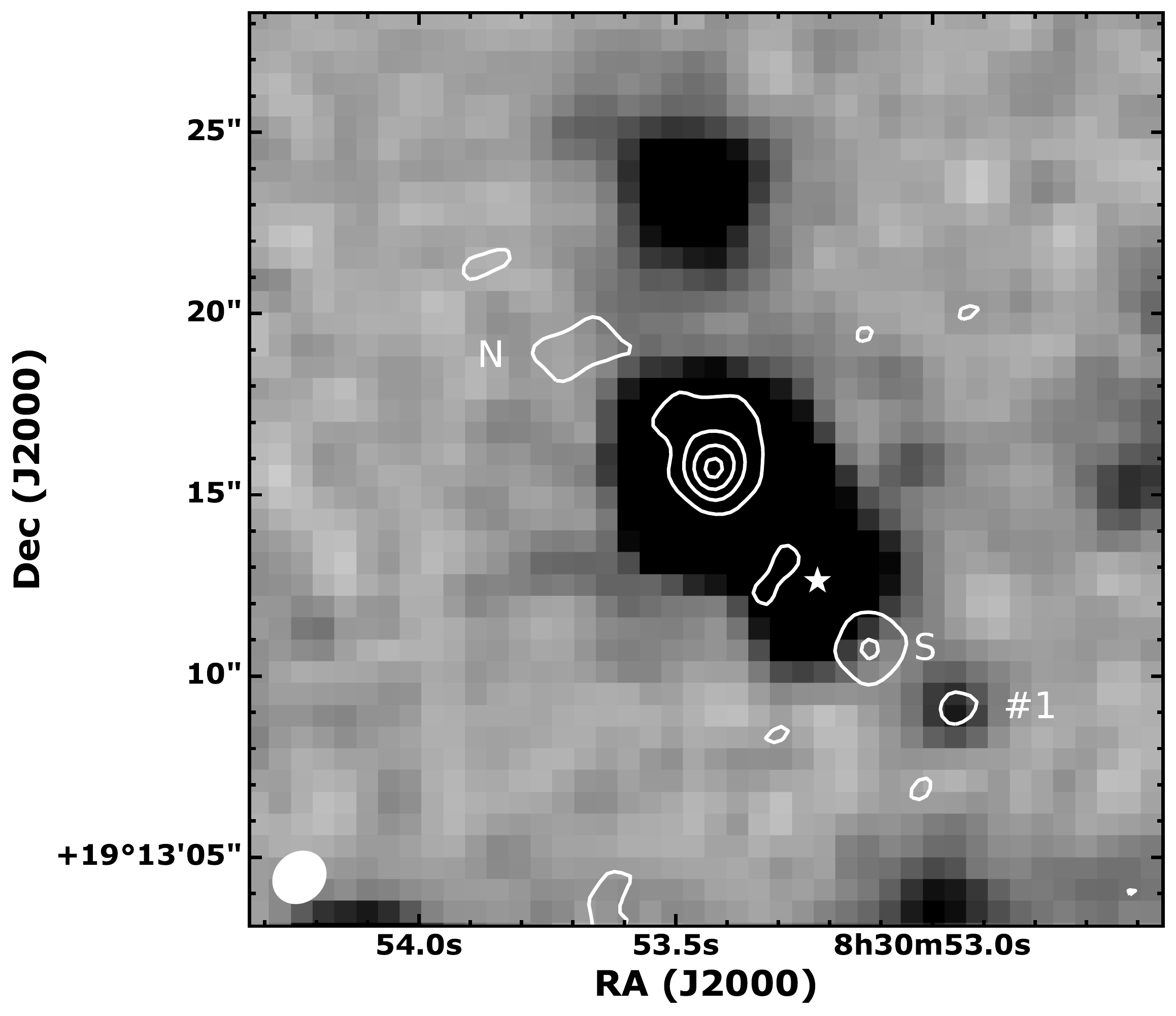}
	\caption{ALMA 100 GHz continuum contours (white) overlayed on the \textit{Spitzer} IRAC $4.5\,\mu\rm m$ image of TXS 0828+193. The contour levels are $3\sigma$, $9\sigma$, $15\sigma$, $21\sigma$, $27\sigma$. The northern and southern hotspots of the radio jets are indicated by letter N and S, respectively. We also indicate the position of companion \#1. The near-IR source south-west to the HzRG, marked with $\star$ symbol, is a known star. The ALMA beam is shown as a white ellipse in the bottom left corner.}
	\label{fig:txs_cont}
\end{figure}

In the ALMA 100 GHz continuum image, TXS 0828+193 is the brightest source with its prominent hotspots detected at the directions of the jets (Fig. \ref{fig:txs_cont}). In addition, a weak continuum emission region, denoted by companion \#1, is marginally detected $\sim$10 arcsec south-west from the HzRG and aligned with the radio jet. The position of companion \#1 coincides with that of the CO emission region reported by \citet{2009MNRAS.395L..16N} and therefore could be thermal dust emission. 
However, given the low flux density of companion \#1 and its alignment with the radio axis, the contamination of the continuum emission by the southern radio jet cannot be excluded. 
The measured continuum flux densities are presented in Table \ref{tab:table1}.

\begin{table*}
	\centering
	\caption{Positions, continuum fluxes and line properties of TXS 0828+193 and its companion candidates.}
	\label{tab:table1}
	\begin{tabular}{lcccccccc}
		\hline
		\hline
		Component &  \multicolumn{2}{c}{Position} & $S_{100\,\rm GHz}$\textsuperscript{a} & $SdV_{\rm CO(3-2)}$\textsuperscript{b} & FWHM & $z_{\rm CO(3-2)}$ & $L_{\rm CO(3-2)}^{'}$ &$M_{\rm H_{2}}$\textsuperscript{c} \\
		 & RA & Dec & $\mu\rm Jy$ & $\rm Jy\,km\,s^{-1}$ & $\rm km\,s^{-1}$ & & $10^{10}\,\rm K\,km\,s^{-1}\,pc^{2}$ & $10^{10}\,\rm M_{\odot}$\\
		\hline
		HzRG & 08:30:53.42 & +19:13:15.71 &  $207\pm25$ & $0.32\pm0.10$ & $616\pm138$ &  $2.5766\pm0.0007$& $1.1\pm0.3$ & $0.9\pm0.3$ \\
		North jet & & & $64.6\pm7.9$ &--& --&  --&--&--\\
		South jet & & & $69.3\pm3.1$ &--& --&  --&--&--\\
		Comp. \#1 & 08:30:52.96 & +19:13:09.09 & $22.5\pm2.1$ & $0.23\pm0.07$ & $635\pm150$ & $2.5727\pm0.0008$ & $0.8\pm0.2$ & $1.0\pm0.3$\\
		Comp. \#2 & 08:30:54.00 & +19:12:47.02 & <44.1 & $0.30\pm0.08$ & $432\pm83$ & $2.5718\pm0.0007$ & $1.1\pm0.3$ & $1.3\pm0.3$\\
		Comp. \#3  (1) & 08:30:54.87 & +19:13:21.67 & <45.9& $0.14\pm0.06$ & $153\pm46$ & $2.5730\pm0.0002$ & $0.5\pm0.$ & $0.6\pm0.3$\\
		Comp. \#3 (2) & 08:30:54.87 & +19:13:21.67 & <45.9& $0.12\pm0.07$ & $287\pm134$ & $2.5682\pm0.0005$ & $0.4\pm0.2$ & $0.5\pm0.3$\\
		Comp. \#4 & 08:30:55.04 & +19:12:56.35 & <51.6 & $0.18\pm0.07$ & $231\pm67$ & $2.5741\pm0.0003$ & $0.6\pm0.2$ & $0.7\pm0.3$\\
		\hline
	\end{tabular}
\flushleft
\footnotesize{$^a$ In the case of non-detection $3\sigma$ upper limits are given, obtained from the primary beam corrected image cube, hence the rms values around the sources are different.}\\
\footnotesize{$^b$ The error on the velocity integrated line flux is from a single Gaussian fit.}\\
\footnotesize{$c$ A conversion factor of $0.8\,\rm M_{\odot}\,(\rm K\,km\,s^{-1}\,pc^{2})^{-1}$ was used to calculate the molecur gas mass. We note that by using a conversion factor of $\sim4\,\rm M_{\odot}\,(\rm K\,km\,s^{-1}\,pc^{2})^{-1}$ found for molecular clouds in the Milky Way and for normal, high-$z$ star-forming galaxies, the molecular mass estimates of the companions are a factor of five higher \citep{2010ApJ...713..686D, 2010MNRAS.407.2091G}.}
\end{table*}

\subsection{CO(3--2) line emission in TXS 0828+193 and companion \#1}
\begin{figure*}
	\centering
	\includegraphics[width=0.8\columnwidth]{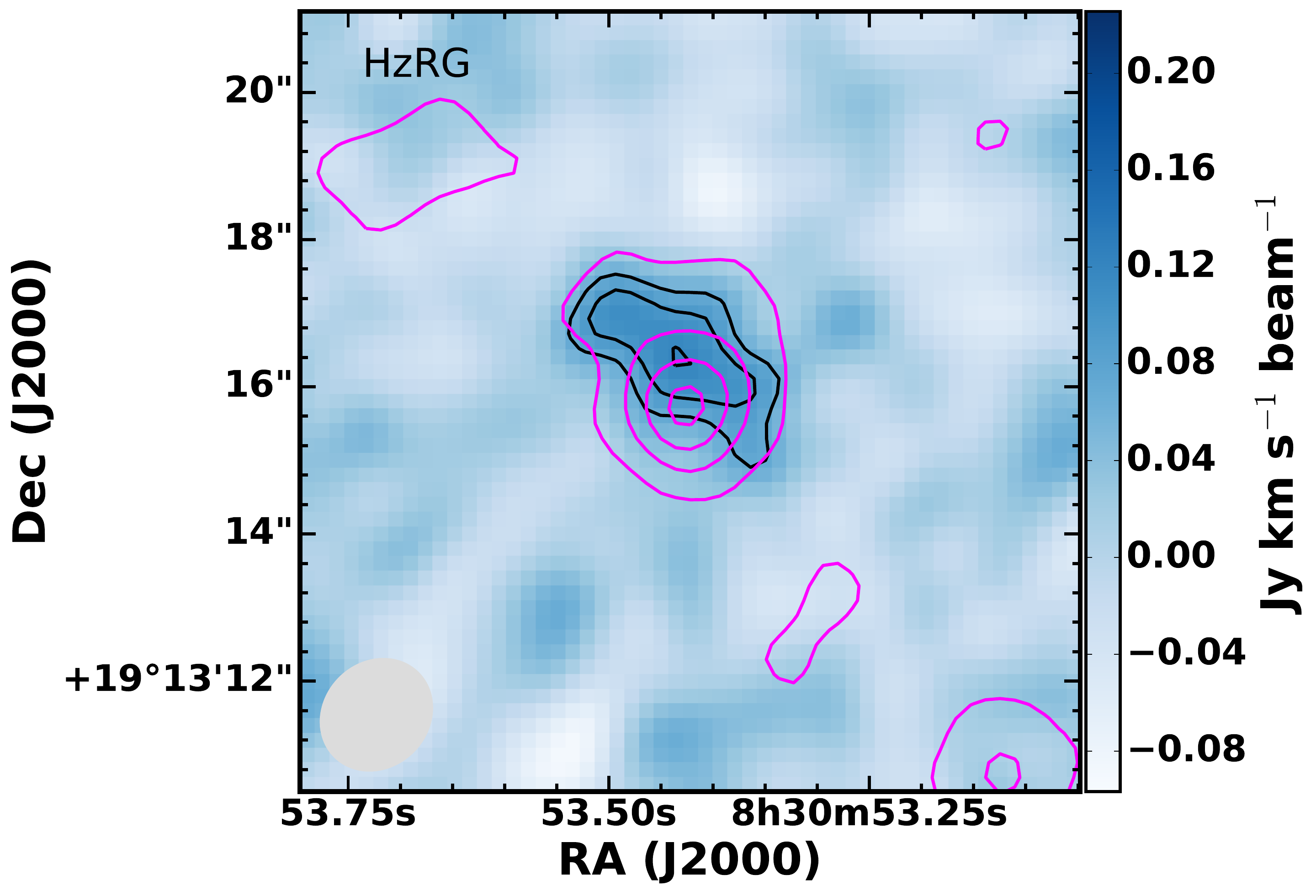}
	\includegraphics[width=0.8\columnwidth]{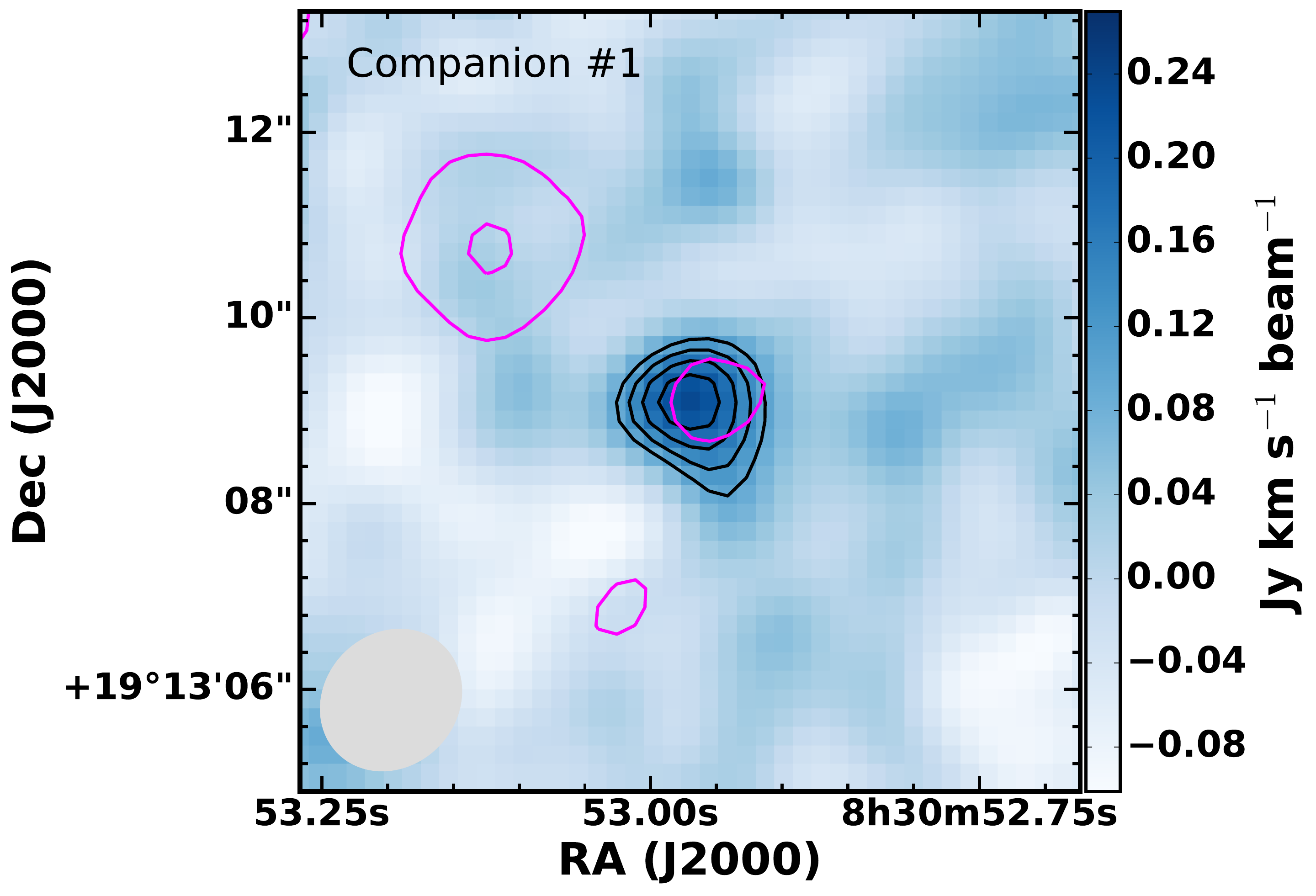}
	\includegraphics[width=0.8\columnwidth]{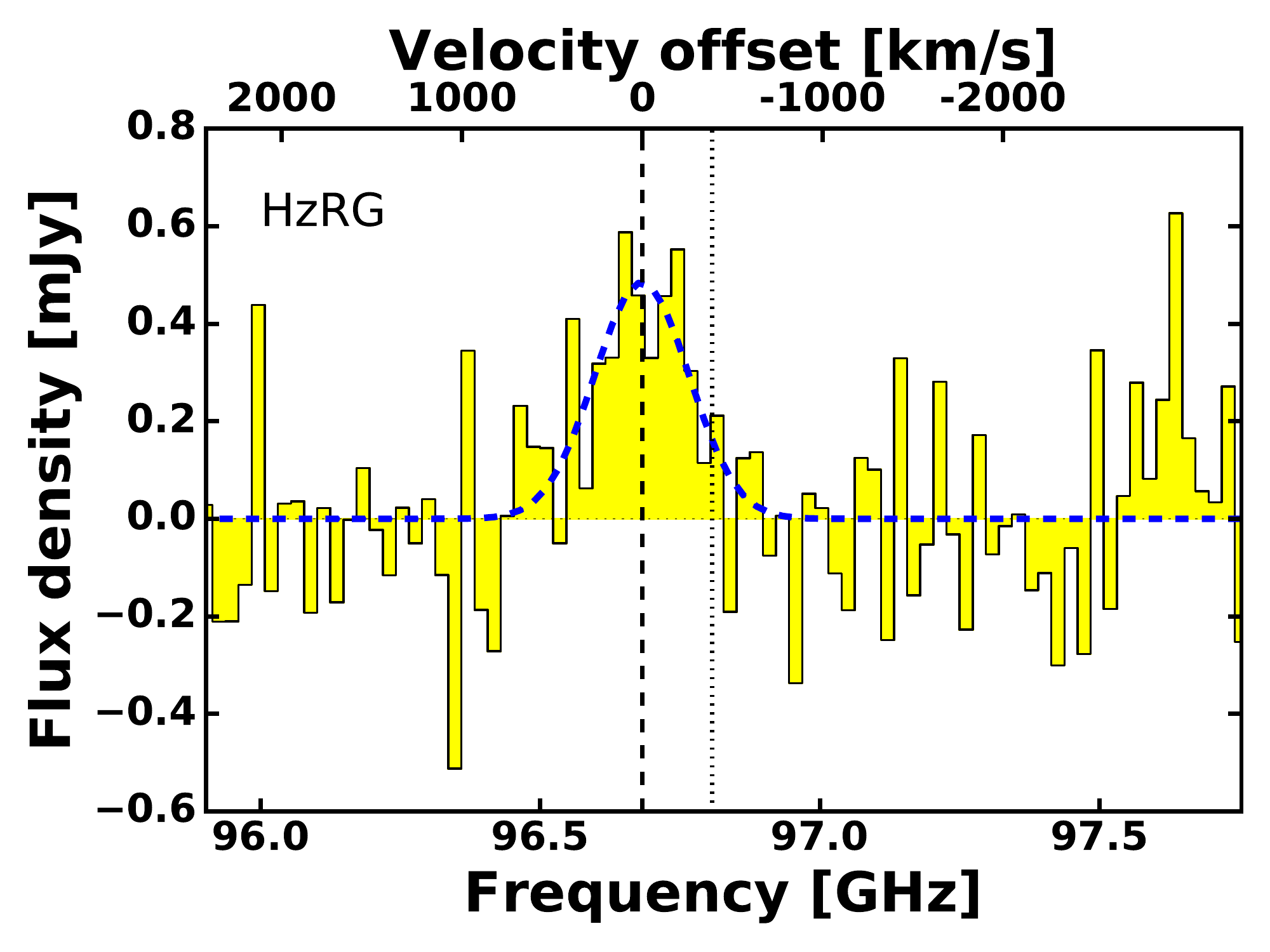}
	\includegraphics[width=0.8\columnwidth]{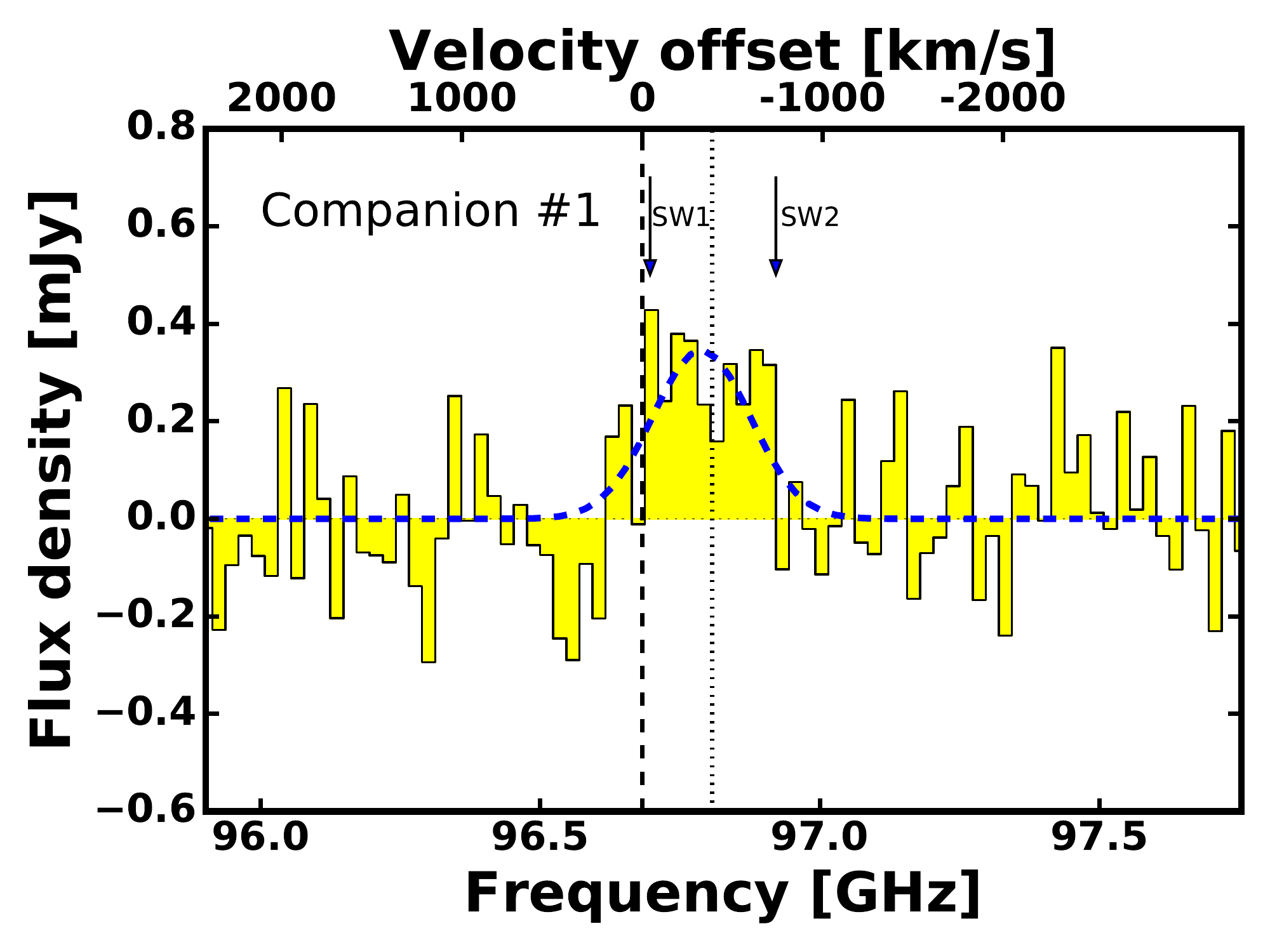}
	\caption{CO(3--2) emission line of TXS 0828+193 and companion \#1. \textit{Top}: The integrated intensity (moment-0) of the HzRG and companion \#1. The moment-0 maps are integrated over the velocity range of $[-394, 319]\,\rm km\,s^{-1}$ ([96.58, 96.81] GHz) in the case of the radio galaxy, and $[-766,225] \,\rm km\,s^{-1}$ ([96.61, 96.93] GHz) in the case of companion \#1, with respect to the CO redshift of the HzRG. The contours start at $3\sigma$ in steps of $1\sigma$, where $\sigma=0.023\,\rm Jy\,beam^{-1}\,km\,s^{-1}$ for the radio galaxy, and $\sigma=0.034\,\rm Jy\,beam^{-1}\,km\,s^{-1}$ for companion \#1. The beam size is indicated as a grey ellipse in the bottom left corner. The magenta contours trace the continuum emission, with the contour levels same as in Fig. \ref{fig:txs_cont}. \textit{Bottom}: The CO(3--2) line spectra are binned to $72.6\,\rm km\,s^{-1}$ per channel and the continuum is subtracted. For each spectrum the whole covered frequency range is shown. The blue dashed curves show the Gaussian fits to the line profiles. The vertical dashed line indicates the CO(3--2) redshift of TXS 0828+193. The vertical dotted line indicates the redshift of the HzRG obtained from $\rm Ly\alpha$ spectroscopic observation ($z=2.572\pm0.002$; \citet{1997A&A...326..505R}). The blue arrows indicate the peak positions of the IRAM PdBI CO(3--2) detection, denoted as SW1 and SW2 in \citet{2009MNRAS.395L..16N}. The top-axis show the relative velocity offset with respect to the CO(3--2) redshift of the HzRG.}
\label{fig:RG_comp1}
\end{figure*}

We detect bright CO(3--2) line emission at the position of the radio galaxy, in contrast to the previous IRAM PdBI CO(3--2) observation \citep{2009MNRAS.395L..16N}. The line emitting region of the radio galaxy is elongated in the NE-SW direction, coinciding with the alignment of the radio lobes (Fig. \ref{fig:RG_comp1}). In addition, our ALMA observation confirm the presence of the offset CO emitting region (companion \#1) first discovered by \citet{2009MNRAS.395L..16N}. However, we only find one unresolved component at the position of companion \#1. The ALMA detected CO(3--2) emission falls right in the middle of the double-peaked profile detected with the IRAM PdBI. The peak positions of SW1 and SW2 are indicated with arrows on Fig. \ref{fig:RG_comp1}.
As the spatial resolution of our ALMA observations is higher than that of the IRAM observation, it is possible that we resolved out the CO emitting region of companion \#1. In order to test this option, we tapered the ALMA data to the beam of the PdBI data ($5.3\arcsec\times4.6\arcsec$). However, tapering the data did not result in detections at higher velocities. The rms of the tapered ALMA data is $\sim0.25\,\rm mJy\,beam^{-1}$ at the centre of the primary beam, with a channel width of $72.6\,\rm km\,s^{-1}$. As the resulted rms is worse than that of the IRAM PdBI observations ($\sim 0.3\,\rm mJy\,beam^{-1}$ per $30\,\rm km\,s^{-1}$ channel), the tapered ALMA data cannot confirm the previously detected double-peak feature of the CO(3-2) emission.

To measure the line properties of the HzRG and companion \#1 and extract their 1D spectra, we first made integrated intensity maps of the sources. After calculating the rms of the moment-0 maps, we defined regions around the sources to include line emission starting at $2\sigma$ level and used them to extract the spectra from the continuum-subtracted data cube. 
We fitted the spectrum with a Gaussian line profile.
Fitting a single Gaussian line to the CO(3--2) emission of the HzRG yields a FWHM of $616\pm138\,\rm km\,s^{-1}$ and a velocity integrated line flux $0.32\pm0.10\,\rm Jy\,km\,s^{-1}$.
The CO emission of companion \#1 has a FWHM of $635\pm150\,\rm km\,s^{-1}$ and a velocity integrated line flux of $0.23\pm0.07\,\rm Jy\,km\,s^{-1}$.
The line profiles and the integrated intensity maps of TXS 0828+193 and companion \#1 are shown on Figure \ref{fig:RG_comp1}.
\subsection{CO(3--2) source finding}
As the radio galaxy is expected to inhabit an overdense environment, the line data cube 
was thoroughly examined in order to find additional CO emitting sources in the field (the $\sim$60 arcsec primary beam corresponds to $\sim$490 kpc at $z=2.572$). For the companion search we used \textsc{SoFiA}, a source finder application suitable for detection and parametrisation of sources in 3D spectral line data cubes \citep{2015MNRAS.448.1922S}. \textsc{SoFiA} has several modules which allow the user to modify the input data cube including flags, filters, weights; to detect spectral lines using three different source finding algorithms and test their reliability; to parametrise sources; to produce output source catalogues, cubes, moment maps of the entire field and moment map cutouts for the detected sources.

We ran \textsc{SoFiA} on the cleaned, natural weighted data cube without applying primary beam correction. For filtering the data we chose the noise scaling option using the standard deviation as a measure of the local noise level. For the source finding setting we used the smooth + clip option with a median absolute deviation rms mode and $5\sigma$ threshold. This means that \textsc{SoFiA} selects sources with a signal-to-noise ratio of $\ge5$.
In the case of the merging option we used the default settings and we did not use parametrisation of the sources. We also tested the source finding including the smoothing option, but this did not change the results significantly.
For a detailed description of the different settings and parameters see \citet{2015MNRAS.448.1922S} and the \textsc{SoFiA} User manual\footnote{\url{https://github.com/SoFiA-Admin/SoFiA/wiki/Documents/SoFiA_User_Manual.pdf}}.

Using the $5\sigma$ threshold \textsc{SoFiA} detected five sources in the frequency range excluding the edge channels, including the HzRG and companion \#1. One of the three new sources found by \textsc{SoFiA} is only a few pixels from the edge of the data cube, thus we exclude it from the further analysis. The other two sources, hereafter companion \#2 and \#3, are located at projected distances of $29.9\,\rm arcsec$ ($\sim245\,\rm kpc$) and $21.6\,\rm arcsec$ ($\sim177\,\rm kpc$) from the HzRG. 

To test the \textsc{SoFiA} results, we also ran the source finding algorithm on the inverted map with a similar setting. This yielded one source at $-5\sigma$ threshold, which suggests that one of the three new sources could be spurious. We note, that by increasing the threshold to $5.5\sigma$ \textsc{SoFiA} only detects the HzRG, companion \#1 and \#2.

\begin{figure*}
	\centering
	\includegraphics[width=0.8\textwidth]{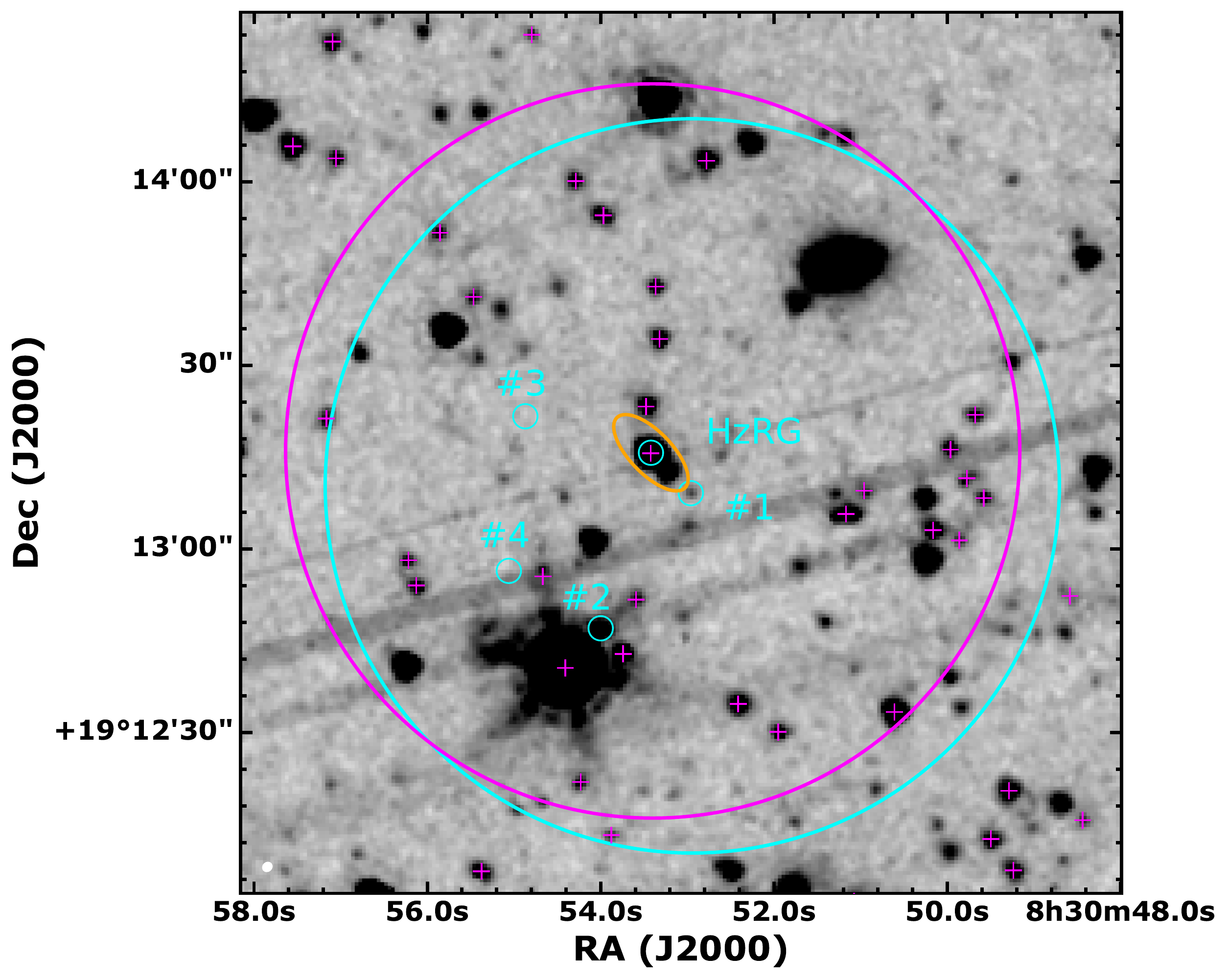}
	\caption{The large-scale environment of TXS 0828+193. The grey-scale image shows the \textit{Spitzer} IRAC $4.5\,\mu\rm m$ map of the field. The large cyan circle indicates the 1 arcmin ALMA primary beam. The small cyan circles mark the position of the HzRG and the CO companions reported in this paper. The large magenta circle shows the 1 arcmin search radius of CARLA, with small magenta crosses marking the positions of the possible cluster members \citep{2013ApJ...769...79W}. The orange ellipse indicates the size of the Ly$\alpha$ halo surrounding the HzRG ($6.9\arcsec\times 16\arcsec$; \citealt{2002MNRAS.336..436V, 2017MNRAS.465.2698M}).}
	\label{fig:txs_comps}
\end{figure*}

We also visually inspected the data cube by collapsing line maps and found one additional source at a projected distance of $30.5\,\rm arcsec$ ($\sim250\,\rm kpc$) from the HzRG, hereafter companion \#4. 
While companion \#4 is only detected by \textsc{SoFiA} at a threshold of $3.5-4\sigma$, it shows clear line emission. The reason to this discrepancy could lie in the way \textsc{SoFiA} estimates the rms noise in the cube and assumes a uniform noise level across it. However, even by setting the local rms noise measurement option in \textsc{SoFiA} did not change the result. The position of the detected companions are marked on Figure \ref{fig:txs_comps}.

To extract the spectra and measure the line properties of the detected companions, we used the same method as for the HzRG and companion \#1, but using the primary beam corrected data cube instead. The velocity integrated line fluxes and line widths are presented in Table \ref{tab:table1}.

\subsection{CO line luminosities and gas masses}
Using the velocity integrated line fluxes of the sources, we estimated their molecular gas mass: in the case of the radio galaxy we use a luminosity ratio of $L'_{\rm CO(3-2)}/L'_{\rm CO(1-0)}=r_{32}=0.97$, while for the companions we use $r_{32}=0.66$, based on the value given for quasars and submillimetre galaxies in \citet{2013ARA&A..51..105C}. After calculating the  $L'_{\rm CO(1-0)}$ luminosities of the sources, we assumed a $\rm CO-H_{2}$ conversion factor of $0.8\,\rm M_{\odot}\,(\rm K\,km\,s^{-1}\,pc^{2})^{-1}$, commonly used for ultraluminous infrared galaxies and starburst galaxies \citep{2005ARA&A..43..677S}. The resulting molecular gas mass estimates are listed in Table \ref{tab:table1}.
We note that in the case of the companions by using a conversion factor of $\sim4\,\rm M_{\odot}\,(\rm K\,km\,s^{-1}\,pc^{2})^{-1}$, found for molecular clouds in the Milky Way and for normal, high-$z$ star-forming galaxies, the molecular mass estimates are a factor of five higher than in Table \ref{tab:table1} \citep{2010ApJ...713..686D, 2010MNRAS.407.2091G}.

\begin{figure*}
	\centering
	\includegraphics[width=0.68\columnwidth]{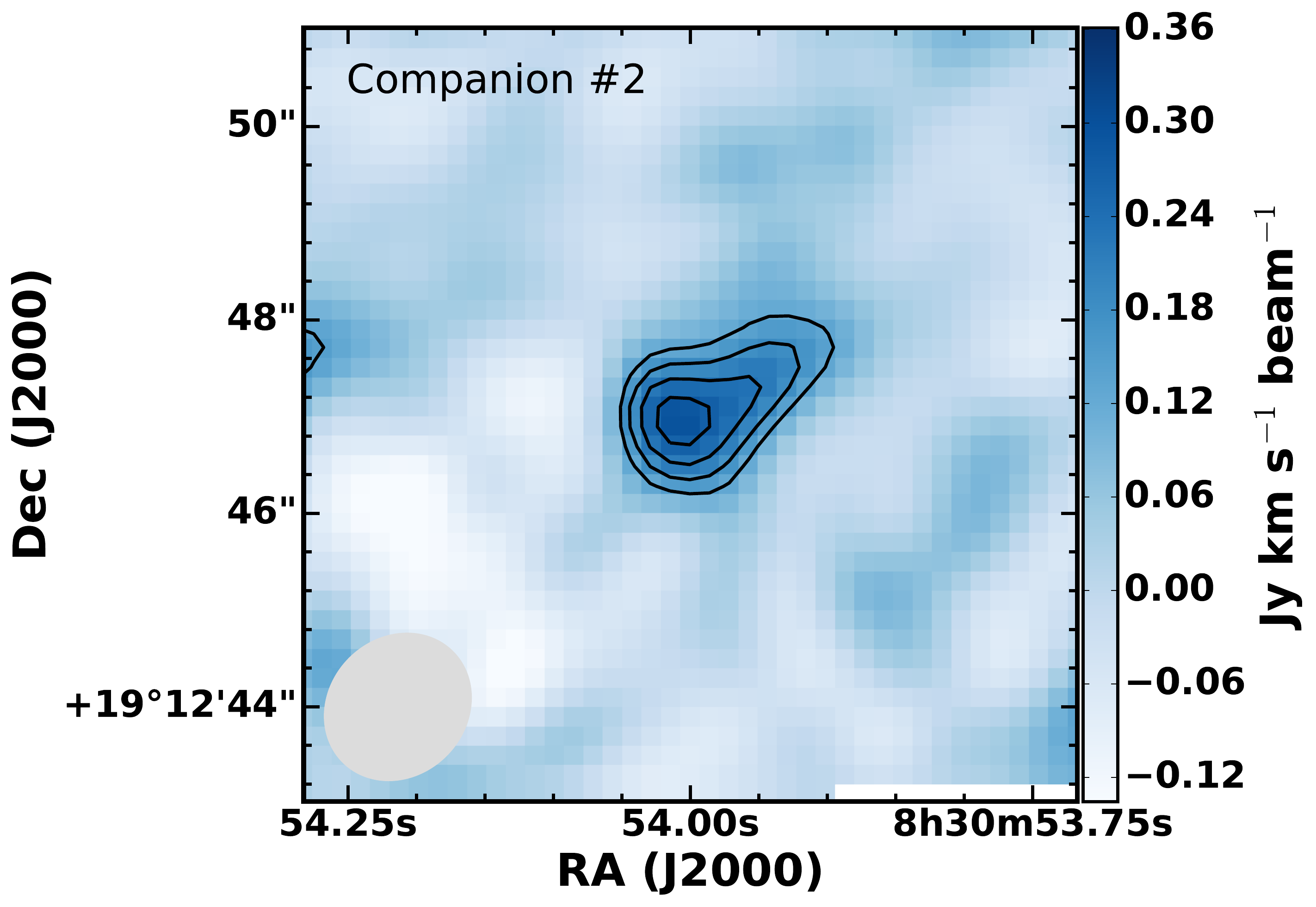}
	\includegraphics[width=0.68\columnwidth]{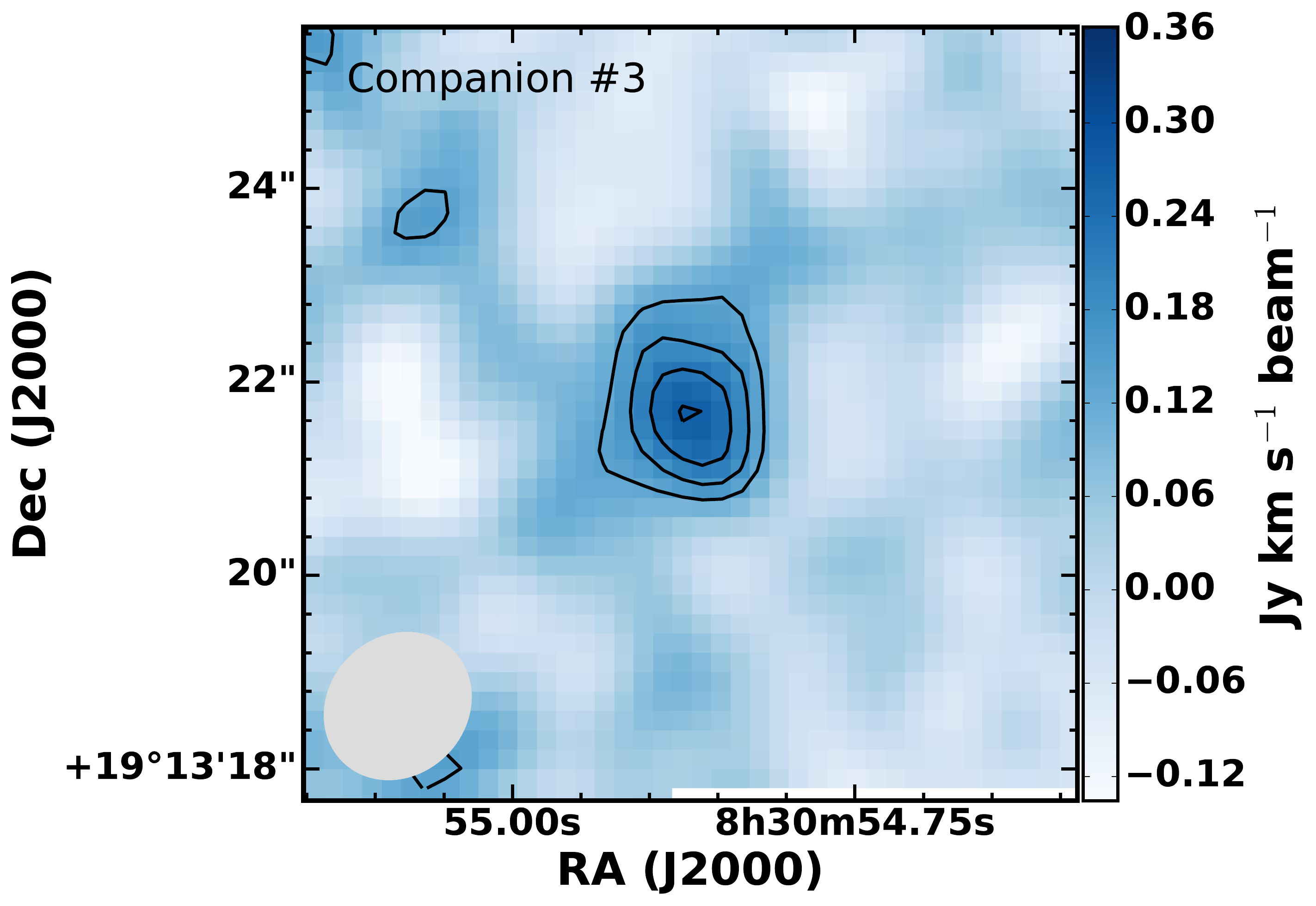}
	\includegraphics[width=0.68\columnwidth]{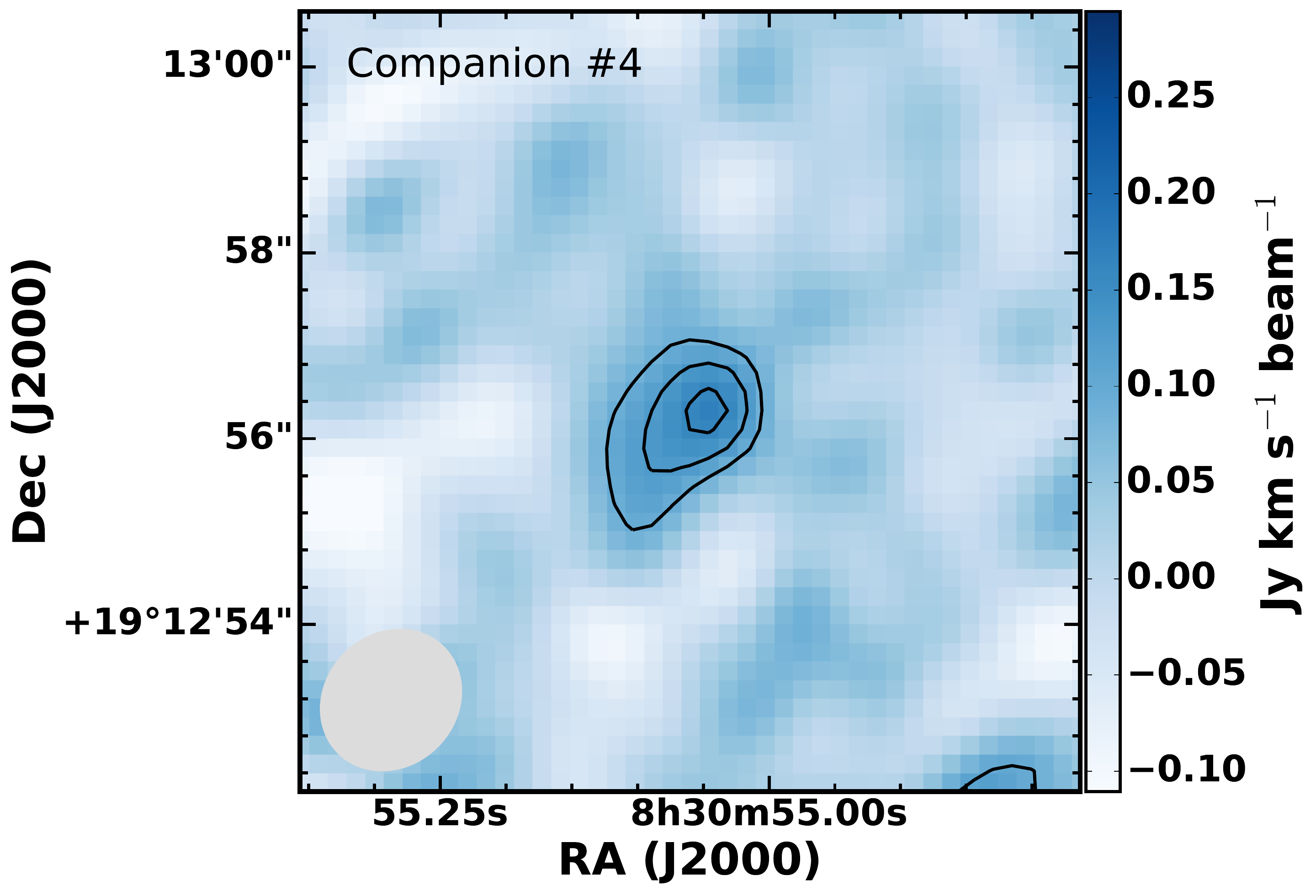}
	\includegraphics[width=0.65\columnwidth]{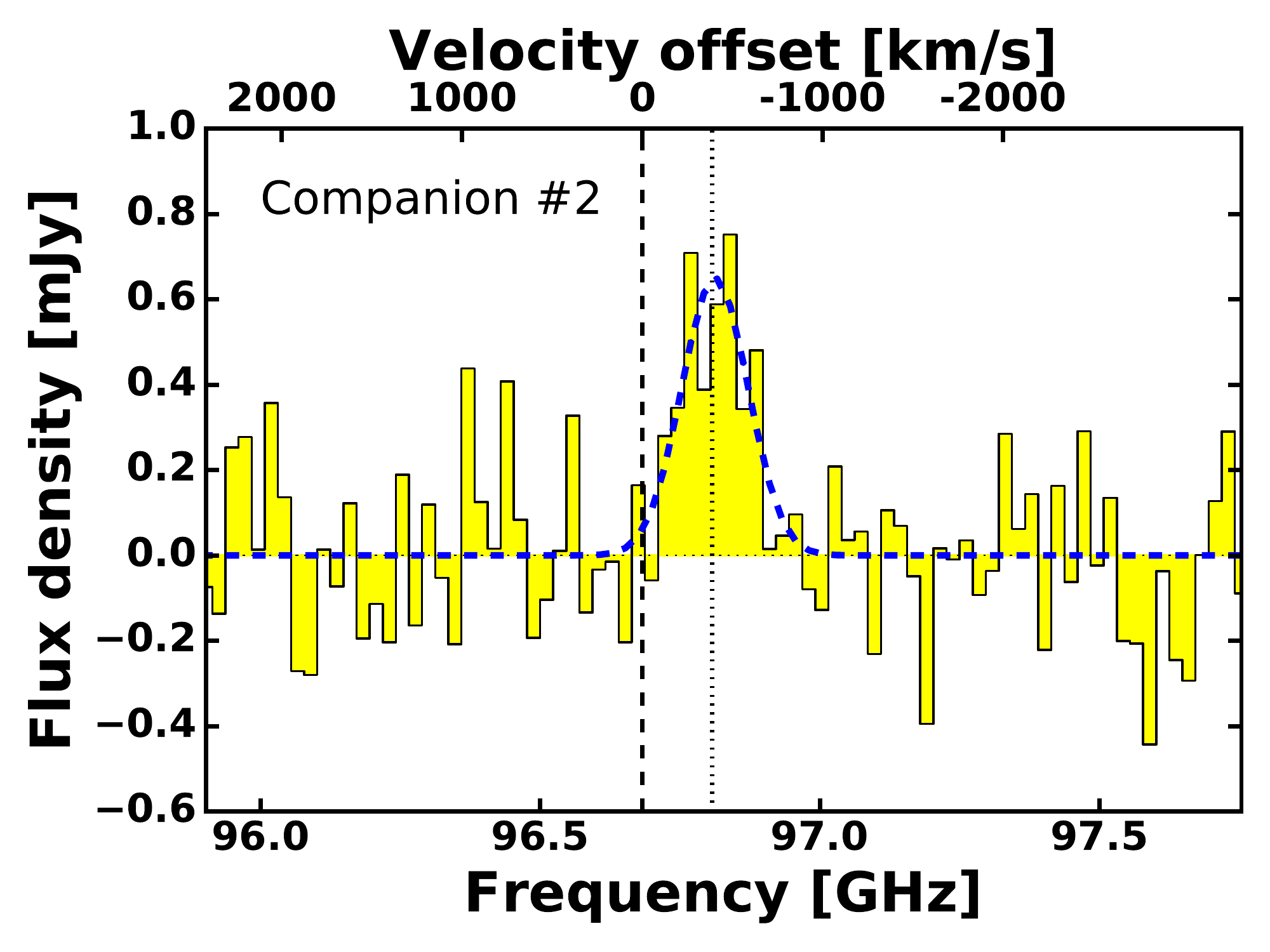}
	\includegraphics[width=0.65\columnwidth]{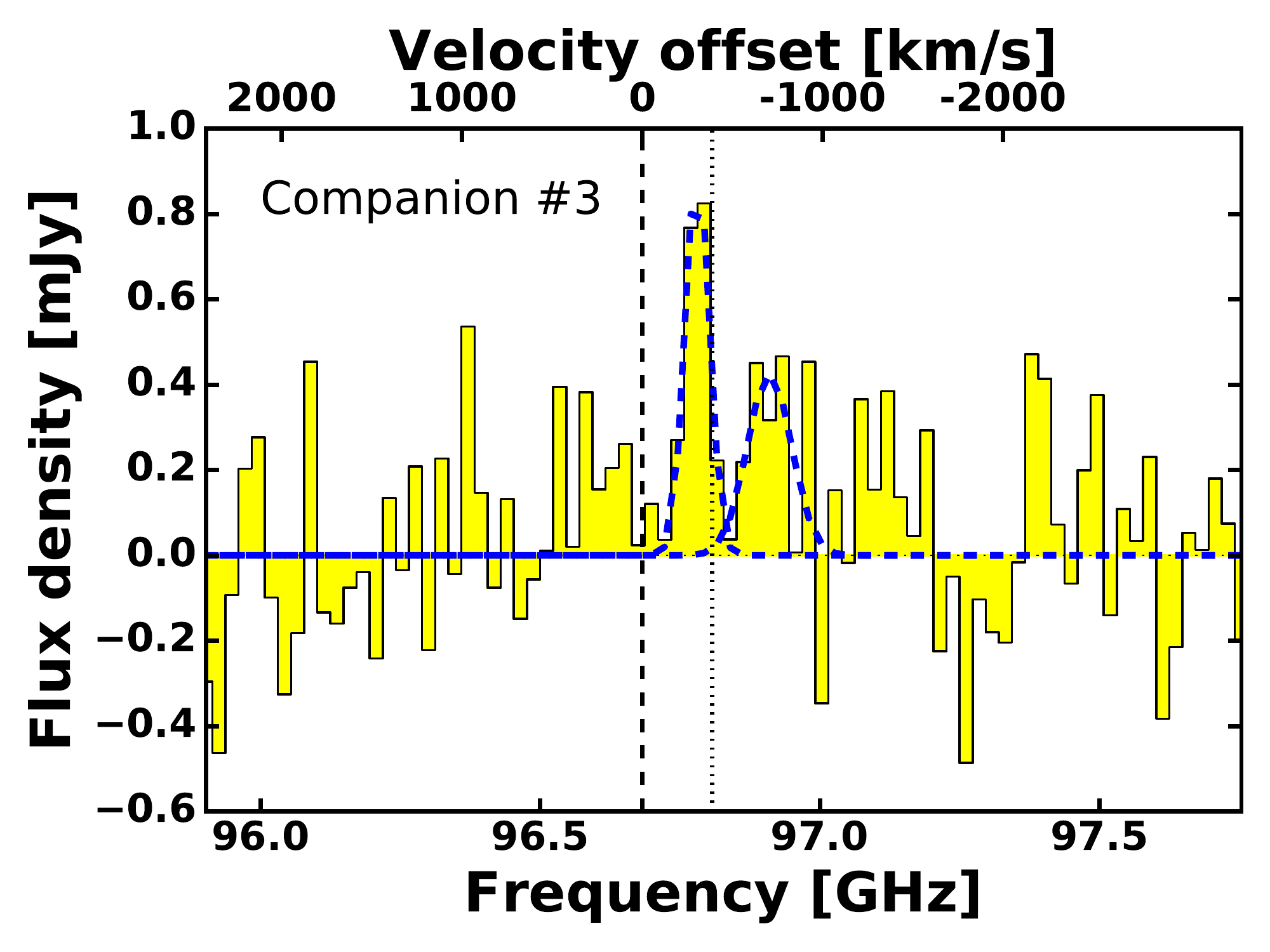}
	\includegraphics[width=0.65\columnwidth]{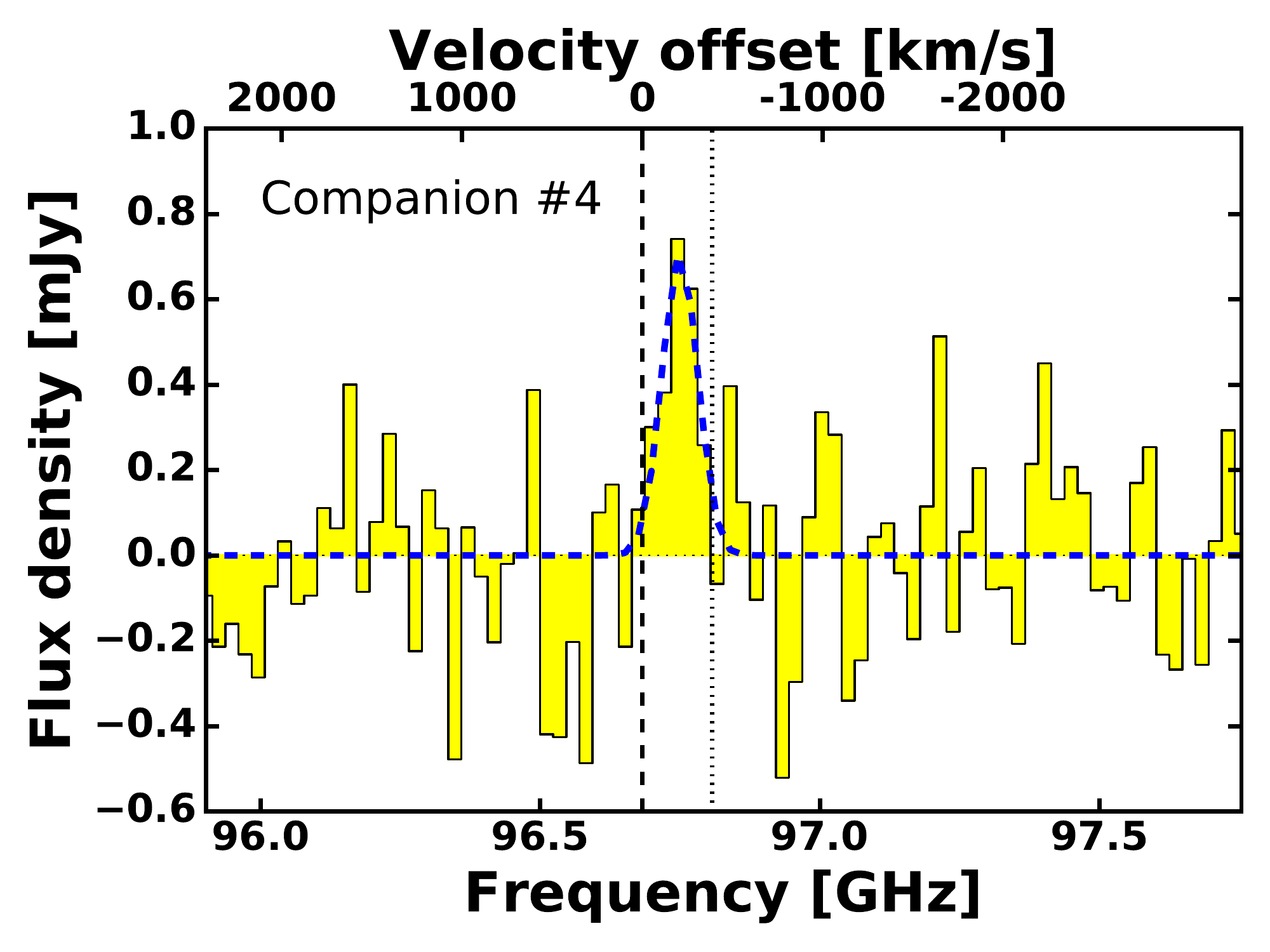}
	\caption{CO(3--2) emission line of the newly detected companions of TXS 0828+193. \textit{Top}: The integrated intensity (moment-0) of the companions. The contours start at $3\sigma$ in steps of $1\sigma$, where $\sigma=0.045\,\rm Jy\,beam^{-1}\,km\,s^{-1}$ for companion \#2 and \#3, and $\sigma=0.034\,\rm Jy\,beam^{-1}\,km\,s^{-1}$ for companion \#4. The beam size is indicated as a grey ellipse in the bottom left corner. \textit{Bottom}: The CO(3--2) line spectra are binned to $72.6\,\rm km\,s^{-1}$ per channel and the continuum is subtracted. In the case of each spectrum the whole covered frequency range is shown. The blue dashed curves show the Gaussian fits to the line profiles. The vertical dashed line indicates the CO(3--2) redshift of the TXS 0828+193. The vertical dotted line indicates the redshift of the HzRG obtained from $\rm Ly\alpha$ spectroscopic observation ($z=2.572\pm0.002$; \citet{1997A&A...326..505R}). The top-axis show the relative velocity offset with respect to the systemic redshift of the HzRG.}
	\label{fig:txs_line}
\end{figure*}
\subsection{Multi-wavelength properties of the companion candidates}
To confirm the presence of the new CO sources we compared the ALMA data to \textit{Spitzer} IRAC archive data. 
The field of TXS 0828+193 has been re-observed by \textit{Spitzer} IRAC at $3.6\,\mu \rm m$ and $4.5\,\mu \rm m$ wavelength (PI: Stern, program ID 80154) as part of the CARLA program \citep{2013ApJ...769...79W}. After careful inspection of the IRAC maps we identified a counterpart of companion \#1, which is more prominent in the $4.5\mu \rm m$ image (Fig. \ref{fig:txs_comps}.). The flux densities of companion \#1 at $3.6\,\mu \rm m$ and $4.5\,\mu \rm m$ are $1.37\pm0.24\,\mu\rm Jy$ and $3.40\pm0.19\,\mu\rm Jy$ (for details about the data reduction and photometry see \citealt{2013ApJ...769...79W}). Companion \#2 is very close to a bright foreground star, which impends its detection in the $3.6\,\mu \rm m$ and $4.5\,\mu \rm m$ maps. The other three companions found in CO(3--2) do not have counterparts in the archive IRAC images.

According to \citet{2013ApJ...769...79W} for a source to be real and considered as a cluster member, its flux density needs to be above the completeness limit in both IRAC bands ($3.45\,\mu\rm Jy$ at $3.6\,\mu \rm m$ and $2.55\,\mu\rm Jy$ at $4.5\,\mu\rm m$), and satisfy the colour criterion of $[3.6]-[4.5]>-0.1$.
The flux density of companion \#1 at $3.6\,\mu \rm m$ is lower than is required by the completeness limit but satisifies the other two criteria. However, the 
ALMA detection of CO(3--2) line emission and 100 GHz continuum emission associated with this source strongly suggests that the IRAC counterpart of companion \#1 is real. To estimate the stellar mass of companion \#1 we used the IRAC $3.6\,\mu \rm m$ observation, which corresponds to the rest-frame \textit{H}-band \citep[e.g.][]{2007ApJS..171..353S}. By assuming a mass-to-light ratio of 1, the IRAC $3.6\,\mu \rm m$ luminosity of the source puts an upper limit of  $\sim 4.7\times10^{9}\,\rm M_{\sun}$ on its stellar mass \citep[e.g.][]{2003ApJS..149..289B}. This is in agreement with the stellar mass estimate of \citet{2009MNRAS.395L..16N}. However, we note that a more accurate way to estimate the stellar mass would require higher quality multiband optical and near-IR photometry and fitting its spectral energy distribution.
\section{Discussion}
\label{sec:discussion}
\subsection{Molecular gas in TXS 0828+193}
In contrast to previous IRAM observations, where only companion \#1 was detected, we successfully detect CO emission associated with the host galaxy of TXS 0828+193. While our ALMA observations were done using six--seven times more antennas than the previous IRAM CO(3--2) observations, resulting in better image quality and a factor of 1.7 times better sensitivity ($\sim 0.1\,\rm mJy\,beam^{-1}$ per $72.6\,\rm km\,s^{-1}$ channel versus $\sim 0.3\,\rm mJy\,beam^{-1}$ per $30\,\rm km\,s^{-1}$ channel of the PdBI data), it is not clear why the radio galaxy was not detected by IRAM, given its line flux is almost a factor of two higher than that of companion \#1.
In addition, the CO line profile of companion \#1 looks different in case of the new ALMA observations, well-fitted with a single Gaussian line instead of a double profile as in \citet{2009MNRAS.395L..16N}. 

With a CO line luminosity of $1.1\,\rm K\,km\,s^{-1}\,pc^2$ and molecular gas mass of $\sim10^{10}\,\rm M_{\odot}$ TXS 0828+193 is similar to other HzRGs found in the literature \citep[e.g.][]{2004A&A...419...99G, 2008MNRAS.390.1117I, 2011ApJ...734L..25E, 2014MNRAS.438.2898E, 2016Sci...354.1128E}. We note that the estimated molecular gas mass heavily depends on the conversion factor used in these studies. What is more interesting is that TXS 0828+193 with its companions perfectly complements the list of HzRGs with off-source CO detections \citep[e.g.][]{2013MNRAS.430.3465E, 2014MNRAS.438.2898E, 2015MNRAS.451.1025E, 2015A&A...584A..99E, 2016A&A...586A.124G, 2017A&A...608A..48D}. Many of these CO emitting regions are at similar or lower distance from the HzRGs than companion \#1 and some of them are associated with dust and/or stellar emission. 
Given the reported galaxy overdenties around HzRGs in mid-IR and submm bands, the discovery of such off-source CO emitting regions is not surprising \citep[e.g.][]{2007A&A...461..823V, 2010MNRAS.405..347F, 2011MNRAS.410.1537H, 2011MNRAS.417.1088K, 2012ApJ...749..169G, 2013A&A...559A...2G, 2013ApJ...769...79W, 2016ApJ...830...90N}. Moreover, these observational findings support the latest results of semi-analytical models of galaxy formation, which imply that the environment of HzRGs is denser, they have more companions in their halos, and the fraction of passive galaxies ($\rm sSFR<10^{-10}\,\rm yr^{-1}$) is higher compared to the environment of QSOs with the same stellar mass \citep{2018MNRAS.480.1340I}. These differences are related to the halo mass of the sources, with the HzRGs residing in more massive ones and AGN feedback, which is identified as the main mechanism influencing the build up of the stellar component of HzRGs. 
\subsection{Off-source CO emission: real or spurious detections?}
The ALMA observations not only detect CO emission associated with the radio galaxy and confirm the presence of companion \#1, but also reveal three additional CO emitting regions at $5\sigma$ significance level. These sources might be potential companion galaxies or massive, cold clumps of gas residing in the halo of the HzRG. However, the possibility of these newly found sources being spurious detections cannot be ruled out completly.
 \citet{2012ApJ...753..135K} argues that the $\ge5\sigma$ detection limit should be reconsidered as the spectral bandwidth of the latest interferometric arrays, such as ALMA, has significantly increased. Thus the broader the spectral bandwidth is, the more likely it is to find adjacent spectral channels with $2\sigma-3\sigma$ peaks, which could lead to spurious detections. Besides this, there are other factors which could increase the number of spurious detections, such as systematic calibration uncertainties, uneven sampling of the $u$-$v$ field and correlated noise between adjacent channels \citep{2012ApJ...753..135K}.
Another option to avoid spurious detections is to only consider sources which have counterparts in other observing bands \citep{2012MNRAS.426..258A}. 
The disadvantage of applying such stricter selection criteria is that they exclude smaller, fainter sources (galaxies or clouds), the detection of which might be important to understand the bigger, more detailed picture of galaxy evolution and clustering, especially in the case of HzRGs, the potential progenitors of the massive elliptical galaxies residing in the centre of local galaxy clusters. 

Apart from companion \#1, none of the CO emitting regions have obvious counterparts in the near-infared archive data, which is puzzling if these sources are young, gas-rich star-forming galaxies. 
As companion \#1 is detected in the ALMA continuum image and has a counterpart in the IRAC bands, we consider this source as a real emission feature. In case of the other companions found in our ALMA data one reason for the lack of detected counterparts might be the sensitivity of the available data: if these sources are small, young or even passive galaxies they might not show up at the achieved sensitivity of the IRAC observations. Regarding dust continuum emission, higher frequency band ALMA observations are more suitable to detect small, dusty galaxies.
Alternatively, the detected line emission of these sources does not originate from the CO(3--2) transition but another line, thus the companions could be galaxies at different redshifts. However, given the fitted CO redshift of the companions are very close to that of the HzRG, this scenario is unlikely.

Even if the companions do not have counterparts in any other observing band, they might be real sources considering their CO line luminosities and their molecular gas masses, that are comparable to that of the host of the HzRG. At $z=2-3$ the available gas in the halos of massive galaxies is still substantial, and it might form clumps at higher density regions. Thus, we also consider companion \#2, \#3 and \#4 as real sources. In the next sections we elaborate on the different scenarios regarding the nature of these sources.
\subsection{Off-source CO emission: companion galaxies or cold gas in the halo?}
Companion \#1 is the closest CO emitting region to the HzRG ($\sim80\,\rm kpc$) and is aligned with the axis of the radio jet, $\sim20\,\rm kpc$ from the south-west radio lobe. Several HzRGs have off-source CO emitting regions along the axis of their radio jets \citep[e.g.][]{2005A&A...430L...1D, 2014MNRAS.438.2898E}, indicating that this alignment cannot be mere coincidence and is related to the interaction of the jets and the surrounding medium. While companion \#1 has a counterpart in the IRAC bands, suggesting that it has a stellar component, it is not clear whether this source is a small galaxy or we witness jet-induced star formation. The latter can happen when the radio jet is propagating through a dense medium and triggering star formation in the overpressured cloud \citep{1987ApJ...321L..29M, 1989ApJ...345L..21B, 1989MNRAS.239P...1R, 1989ApJ...342L..59D}. Such jet-induced star formation has been observed both at low and high redshift \citep[e.g.][]{2004ApJ...612L..97K,2004IAUS..217..472V, 2007ApJ...662..872F, 2008MNRAS.386.1797I, 2008A&ARv..15...67M, 2015IAUS..309..182C}. 

Long-slit spectroscopic observations of TXS 0828+193 revealed a large Ly$\alpha$ halo, extending $\sim$16$\,\rm arcsec$($\sim$130 kpc) along the radio axis and $\sim6.9\,\rm arcsec$ ($\sim$56 kpc) perpendicular to it \citep{2002MNRAS.336..436V, 2017MNRAS.465.2698M}. The inner  $\sim$30 kpc of the halo gas is kinematically perturbed along both slit positions, indicating the presence of a bubble of expanding gas powered by feedback processes from the central radio galaxy \citep{2017MNRAS.465.2698M}. The outer part of the Ly$\alpha$ halo seems to be kinematically unperturbed. Companion \#1 is located close to the edge of the Ly$\alpha$ halo but far from the outflow when comparing the recent data from \citet{2017MNRAS.465.2698M}. From the available data it is not clear if companion\#1 is related to the Ly$\alpha$ halo but it is possible that the CO emission originated from the diffuse halo gas and was later compressed by the radio jet triggering star formation.

Companion \#2, \#3 and \#4 are only detected in CO emission, thus for these sources to be galaxies, they need to be in a very young stage of their stellar build up in order to explain their non-detection in the IRAC bands (for more detailed discussion on this scenario we refer to reader to Section 3.1 of \citealt{2009MNRAS.395L..16N}). Moreover, these sources are at greater distances from the HzRG than companion \#1 and are not aligned with the radio axis, thus ruling out the possibility of jet-induced gas cooling or jet-driven enrichment of the intergalactic medium (IGM; see \citealt{2014MNRAS.438.2898E} and the references therein).
A more plausible scenario regarding the nature of these sources is that they are massive gas clouds in the halo, along the large-scale filamentary structure of the cosmic web. According to cosmological simulations the gas of the IGM is distributed in filaments and sheets, especially at high-redshift ($z\sim3$), where the majority of gas is concentrated in such structures \citep{2006Natur.440.1137S, 2019MNRAS.486.3766M}. Galaxies, groups and clusters form and evolve in the intersections of the filaments, which provide the necessary gas for their growth in the form of cold streams \citep[e.g.][]{2009Natur.457..451D, 2009ApJ...703..785D, 2011MNRAS.418.1796F, 2012MNRAS.423..344R}. While the gas in the filaments has low density, which makes it difficult to observe, there has been some interesting results in the recent years in this front using absorption spectroscopy to trace neutral hydrogen, searching for $\rm Ly\alpha$ emission around star-forming galaxies and AGN or using $\rm Ly\alpha$-emitting galaxies as tracers \citep[e.g.][]{2014Natur.506...63C, 2015Natur.524..192M, 2015MNRAS.449.1298S, 2016ApJ...831..181L, 2016ApJ...831...39B,2017A&A...602L...6V, 2019Sci...366...97U}. One particulary interesting study targeted a $z=3.09$ protocluster using MUSE on VLT and found extended $\rm Ly\alpha$ emission, outlining a filamentary structure across the observed field, with brighter emission regions associated with the circumgalactic medium of galaxies and low-surface brightness emission connecting to individual galaxies \citep{2019Sci...366...97U}. 
Based on these results, the CO detected companions of TXS 0828+193 might be high-density regions of the diffuse gas in the halo of the HzRG and potentially linked to large-scale filamentary structures. 
\section{Conclusions}
We have studied the $z=2.57$ radio galaxy TXS 0828+193 and its environment using ALMA CO(3--2) observations. We detected CO emission associated with the host galaxy of the HzRG ($M_{\rm H_{2}}=(0.9\pm0.3)\times10^{10}\,\rm M_{\odot}$), in contrast to previous interferometric observations. We also confirmed the presence of the off-source CO emitting region (companion \#1) discovered by \citet{2009MNRAS.395L..16N}. As this source has a counterpart in our continuum data and in the \textit{Spitzer} IRAC bands, it could be either a normal star-forming galaxy in the vicinity of the HzRG or the result of jet-induced star formation, given its apparent alignment with the jet axis of TXS 0828+193.

In addition, we detected three new CO emitting regions in the field of TXS 0828+193 at a distance of $\sim170-250\,\rm kpc$. These companions have comparable molecular gas masses to the HzRG and do not have counterparts in any other observing band. Since these sources are not aligned with the axis of the radio jet and are at higher distances from the HzRG than companion \#1, they are likely cold, dense clouds in the environment of the central HzRG, potentially part of the large-scale filaments of the cosmic web, which provide the necessary gas for the growth of galaxies. 

Based on the available data it is difficult to assess whether these clouds will provide fuel for the AGN or the stellar build up of the HzRG or become galaxies themselves, thus populating a cluster around TXS 0828+193. However, our results together with other recent ALMA results of the HzRG environment demonstrate the potential for progress on our knowledge of early galaxy cluster evolution and the evolution of massive AGN host galaxies through gas observations. The way forward to establish a complete understanding of such environments requires dedicated continuum and line observations.

\label{sec:conclusion}

\section*{Acknowledgements}
We would like to thank the reviewer for the constructive report that helped us to improve the manuscript.
JF acknowledges Dominika Wylezalek for kindly providing the IRAC photometry results of TXS 0828+193.
JF acknowledges support from the Nordic ALMA Regional Centre (ARC) node based at Onsala Space Observatory. The Nordic ARC node is funded through Swedish Research Council grant No 2017-00648.
JF and KK acknowledge support from the Knut and Alice Wallenberg Foundation. KK acknowledges support from the Swedish Research Council.
This paper makes use of the following ALMA data: ADS/JAO.ALMA\#2015.1.00661.S. ALMA is a partnership of ESO (representing its member states), NSF (USA) and NINS (Japan), together with NRC (Canada) and NSC and ASIAA (Taiwan) and KASI (Republic of Korea), in cooperation with the Republic of Chile. The Joint ALMA Observatory is operated by ESO, AUI/NRAO and NAOJ.

\section*{Data availability}
The data underlying this article will be shared on reasonable request to the corresponding author.

\bibliographystyle{mnras}
\bibliography{fogasy_txs0828_final} 
\bsp	
\label{lastpage}
\end{document}